\documentstyle[preprint,aps]{revtex}

\begin{document}

\title{Imaging Sources with Fast and Slow Emission Components}

\author{G. Verde$^{a}$, D.A. Brown$^{b}$, P. Danielewicz$^{a}$, 
C.K. Gelbke$^{a}$, W.G. Lynch$^{a}$, M.B. Tsang$^{a}$}
\author{\small $^{a}$ National Superconducting Cyclotron Laboratory 
and Department of Physics and Astronomy, 
Michigan State University, East Lansing, MI, 48824-1321, USA}
\author{\small $^{b}$Lawrence Livermore National Laboratory, 
Livermore, CA 94551-0808, USA}

\date{\today}

\maketitle

\begin{abstract}
We investigate two-proton correlation functions for reactions 
in which fast dynamical and slow evaporative proton emission 
are both present. In such cases, the width of the correlation 
peak provides the most reliable information about the source 
size of the fast dynamical component. The maximum of the correlation 
function is sensitive to the relative yields from the slow and 
fast emission components. Numerically inverting the correlation 
function allows one to accurately disentangle fast dynamical 
from slow evaporative emission and extract details of the shape 
of the two-proton source.
\end{abstract}

%\pacs{25.70.-z, 25.70.Pq}

\newpage

Intensity interferometry, the investigation of two-particle 
correlation functions at small relative momenta, can provide 
important information about the space-time characteristics and 
underlying dynamics of particle emitting sources [1-6].  At incident 
energies below the pion threshold, two-proton correlation functions 
[1,4,5,7-20] have been the tool of choice for studying various 
equilibrium and non-equilibrium emission processes. Initial analyses 
of two-proton correlation functions in terms of instantaneous 
emission from sources with Gaussian density profiles have yielded 
source radii that strongly depend on the energy of the emitted 
particles [9,12,13,21], indirectly signaling the importance of 
lifetime, expansion, and/or cooling effects.  Later comparisons
of measured correlation functions to predictions of reaction 
models, assuming either slow evaporative emission [11,12,15] 
or fast emission according to BUU transport calculations 
[12,13,16-19,21], have provided more quantitative understanding.  
In particular, some of the correlation functions measured for 
intermediate-energy nucleus-nucleus collisions have been 
quantitatively reproduced by BUU transport calculations in 
a number of cases [12,13,16]. 
Other data, primarily at higher incident energies [17-19], indicated 
emission time scales significantly longer than predicted by BUU 
transport theory. This led to speculations [19] that proton emission 
consists of two major components involving significantly different 
time scales -- a fast component consistent with BUU transport 
theory predictions and a slower component that might originate 
from the sequential decay of unstable nuclei produced in 
multifragmentation reactions [19]. 
Such scenarios may be rather common, with the 
decay of an equilibrated residue contributing to the slow component 
at incident energies below the multifragmentation threshold and 
the decay of excited fragments from multifragmentation processes 
contributing to the slow component at higher incident energies. 
When such slow components are present, quantitative comparisons 
to dynamical models are contingent upon a determination of the 
relative contributions of these fast and slow components [19]. 

Up to now, correlation function analyses have focused upon obtaining 
a satisfactory description of the maximum of the correlation 
function and comparatively little attention was paid to its shape. 
Numerical inversion techniques have been recently developed [22-24] 
that provide the capability to analyze the shape of the correlation 
function and thereby extract much more detailed information about 
the particle emission mechanism. In this paper, we apply these 
techniques to analyze correlation functions for systems with 
strong admixtures of fast and slow emission components. Using
general arguments, we show that the width of the correlation 
peak provides a clear measure of the source size for the fast 
component and that the height of the correlation peak provides 
the key information about the relative yields from the fast and 
slow emission components. We confirm these insights by a detailed 
re-analysis of the shapes of two-proton correlation functions 
that were measured previously for $^{14}$N+$^{197}$Au collisions 
at E/A=75 MeV [12] and were described rather unsatisfactorily [12] 
by both BUU calculations and by zero-lifetime Gaussian sources.

Experimentally, the (angle-integrated) two-proton correlation 
function $1+R(q)$ is defined in terms of the two-particle coincidence 
yield $Y_{2}({\mathbf{p}}_{1},{\mathbf{p}}_{2})$ and the single particle yields 
$Y_{1}({\mathbf{p}}_{1})$ and $Y_{1}({\mathbf{p}}_{2})$:
\begin{equation}
\sum Y_{2}\left({\mathbf{p}}_{1},{\mathbf{p}}_{2}\right) = 
C\cdot\left(1+R(q)\right)\sum 
Y_{1}\left({\mathbf{p}}_{1}\right)Y_{1}\left({\mathbf{p}}_{2}\right)
\label{eq:correlation}
\end{equation}
Here, ${\mathbf{p}}_{1}$ and ${\mathbf{p}}_{2}$ are the laboratory momenta of 
the two coincident particles, $q=\mu\cdot{v_{rel}}$ 
is the momentum of relative motion, and $C$ is a normalization 
constant chosen such that ${\langle}R(q){\rangle}=0$ 
for large $q$ where final-state interaction effects are negligible.  
For a given experimental gating condition, the sums on each side 
of Eq. (1) extend over all particle energies and detector 
combinations corresponding to each bin of $q$.

Theoretical correlation functions are calculated from the 
angle-averaged Koonin-Pratt formula [1,3,10,23]
\begin{equation}
R(q) = 4\pi\int{dr}{\cdot}r^{2}K(q,r){\cdot}S(r)
\label{eq:koopra}
\end{equation}
where the source function $S(r)$ is defined as the probability 
distribution for emitting a pair of protons with relative distance $r$
at the time the second proton is emitted. 
The source function reflects both the spatial extent and lifetime 
relevant to single proton emission. An increase in the emission 
timescale broadens $S(r)$, 
and for typical proton velocities $v_{proton}$ 
and timescales $\tau$ such that $v_{proton}\cdot\tau$ 
is comparable to or larger 
than $r_{0}$, the complications of finite emission timescales need 
special consideration [1,5,10,11,14,15,20-25]. The difficult 
problem posed by the simultaneous presence of very short and 
very long time scale emissions is discussed below. The angle-averaged 
kernel, $K(q,r)$, 
is calculated from the radial part of the antisymmetrized 
two-proton relative wave-function. At short distances, $K(q,r)$
is dominated by the strongly attractive singlet S-wave proton-proton 
interaction which gives rise to a maximum in $R(q)$ at relative 
momentum $q\approx{20}$ MeV/c.  
The long-range repulsive Coulomb interaction and antisymmetrization 
both produce a minimum at $q\approx{0}$ MeV/c [1]. 
Both the peak at $q\approx$ 20 MeV/c and the minimum at $q\approx{0}$ 
MeV/c are clearly visible in the correlation function represented 
by the solid line in the right panel of Fig. 1. 

The solid line in the left panel of the same 
figure represents the source function for this correlation function. 
Its shape, reflecting both short and long-ranged components, 
is representative of intermediate energy heavy ion reactions. 
The short-ranged contribution, approximated by the dashed line 
in the left panel, peaks at $r=0$ and originates from fast dynamical 
emissions that dominate the earlier stages of the reaction. This 
short-range contribution generates the correlation function described 
by the dashed line in the right panel. The long-ranged contribution, 
approximated by long exponential tail at large r-values (dot-dashed 
line, left panel), corresponds to the emission of one or both 
protons via long time scale secondary decays of excited fragments 
or the heavy residue, which are produced in the same collision. 
Unlike the fast component, the secondary decays that contribute 
to the long-ranged contribution occur over much longer time scales 
than can be reliably described by transport theory. Moreover, 
the influence of this long time-scale component (dot-dashed line, 
right panel) is negligible everywhere except at $q<15$ MeV/c, 
where the Coulomb interaction and the Pauli principle are major 
factors and the measurements are very difficult. 

The similarity of the dashed and solid lines at $q>10$ MeV/c in 
Fig. 1 makes it clear that the correlation function, at such 
relative momenta, is sensitive only to the fast dynamical portion 
of this source. However, the long time scale evaporative decays 
cannot be ignored. To illustrate clearly their influence, let 
us discuss an extreme example where emission is dominated by 
two-components, 1) an instantaneous fast component ($\tau_{1}=0$) 
and 2) a slow component with sufficiently long decay timescale that 
its correlation function is negligible ($\tau_{2}\approx\infty$).  
If fast emission provides a fraction $f$ of the emitted protons, 
$Y_{1,fast} = f\cdot{Y_{1}}$, and if slow emission provides the 
remainder, $Y_{1,slow}=(1-f)Y_{1}$, the resulting correlation 
function $R(q)$ will reflect only the fast source and will be given 
by

\begin{equation}
R(q)=f^{2}{\cdot}R_{fast}(q){\equiv}{\lambda\cdot}R_{fast}(q)
\label{eq:fastslow}
\end{equation}
where $R_{fast}(q)$ denotes the correlation function when only fast 
emission is present.  Equation (3) stipulates that the height 
of the correlation function peak, $R(20 MeV/c)$, is attenuated 
by the factor $\lambda=f^{2}$; thus, when fast and slow emission 
processes admix, the height of the correlation peak alone does 
not determine the source size. 
However by its very construction, any property solely derived 
from the shape of the correlation function for fast emission 
such as the width of the correlation peak, ${\Delta}q_{FWHM}$, 
will remain unaffected. 

For simplicity, we assume that the fast component has a simple 
Gaussian profile,
\begin{equation}
S_{G}(r)=\frac{1}{(2\pi )^{3/2}
  r_{0}^{3}}\exp\left(-\frac{r^{2}}{2r_{0}^{2}}\right)
\label{eq:sougauss}
\end{equation}
where ${\sqrt{3}}{\cdot}r_{0}$ is the Gaussian rms relative 
proton-proton source radius.  For such Gaussian sources, 
the parameter $r_{0}$ describes the spatial 
distribution of emission points for zero lifetime sources [1]. 
For $\lambda=f^{2}=1$, $r_{0}$ is uniquely related to the height, 
R(20 MeV/c), (see the curve with $f=1$ in the lower right panel 
of Fig. 2) and to the width, $\Delta{q}_{FWHM}$, 
(see the top right panel of Fig. 2) of the correlation peak.  
This simple two-component model (STCM) thus contains two parameters, 
the Gaussian source radius $r_{0}$ and the fraction of coincident 
pairs resulting from the fast emission components, $\lambda=f^{2}$.
Results obtained with the STCM are illustrated in Fig. 2. 
Consistent with the previous discussion, the top right panel 
demonstrates that $\Delta{q}_{FWHM}$ depends uniquely on the 
radius of the fast source, $r_{0}$. 
The bottom right panel demonstrates that $R(20 MeV/c)$ depends 
on both $r_{0}$ and $f$.  The left panel further emphasizes 
this latter feature by showing that the same value of the correlation 
maximum can be achieved for different values of $r_{0}$ and $f$.  

Thus, measurements of the width, $\Delta{q}_{FWHM}$, 
of the correlation function provide a direct measure of the size of 
the source. Measurements of both $\Delta{q}_{FWHM}$ and the
correlation function maximum, $R(20 MeV/c)$, provide a unique 
determination of the relative strengths of equilibrium 
and dynamical emission components.  This finding should remain 
approximately valid for more realistic emission time scales, 
$0<\tau_{fast}<<\tau_{slow}<\infty$, 
provided $v_{proton}\cdot\tau_{slow}>>r_{source}$ and 
$v_{proton}\cdot\tau_{fast}\approx{r_{source}}$. 
We emphasize that the correlation function is 
only sensitive to the details of the long-lived component at 
very small relative momenta, $q<15$ MeV/c [1,4,5,11,15] where 
the long-lived component (dot-dashed line in Fig. 1) generates 
a non-vanishing correlation function, $R_{slow}(q)\not= 0$. 
It may be difficult to utilize this sensitivity to probe the 
details of the long-lived component when the correlation function 
exhibits a large peak at 20 MeV/c reflecting the additional presence 
of a strong short-lived component. 

We now apply these ideas to the interpretation of two-proton 
correlation functions measured at 
$\theta_{lab}{\approx}$25$^{o}$ 
for $^{14}$N+$^{197}$Au collisions at E/A=75 MeV [12].  
The points on 
the left panel of Fig. 3 show the correlation functions measured 
for proton pairs with three different cuts on their total momentum 
$P_{tot}=\left|{\mathbf{p}}_{1}+{\mathbf{p}}_{2}\right|$, 
270 MeV/c $<P_{tot}<$ 390 MeV/c (circular points), 
450 MeV/c $<P_{tot}<$ 780 MeV/c (triangular points), 
and 840 MeV/c $<P_{tot}<$ 1230 MeV/c (square points).  
As observed for many other reactions [9,12,13,18,19,21], 
the height of the correlation peak increases with increasing
$P_{tot}$.  
Correlation functions for zero-lifetime Gaussian sources are shown 
as dot-dashed curves; the measured values of $R(20 MeV/c)$ can be 
reproduced by using strongly momentum dependent radius parameters 
$r_{0}=5.9$ fm, $4.2$ fm, and $3.4$ fm for the low, medium and high 
momentum gates, respectively, but the widths of the correlation 
function peaks cannot be reproduced by these calculations [12]. 
Indeed, the widths of the measured correlation functions remain 
roughly constant at about $\Delta{q}_{FWHM}\approx{19-22}$ MeV/c, 
which implies nearly constant source radii according to Fig. 2. 
This inconsistency means that these strongly momentum-dependent 
radius parameters are incorrect. 

A somewhat better description of this experimental data can be 
attained with the STCM by suitable choice of its free parameters 
($r_{0}$ and $\lambda$ in Eq. (3) and (4)) [26]. 
The STCM approach describes well data  that are dominated by 
fast dynamical proton emission; the STCM fit for the high momentum 
gate 840 MeV/c $<P_{tot}<$ 1230 MeV/c is virtually identical to the 
fit obtained by the numerical inversion procedure discussed below. 
(The results of the numerical inversion procedure are shown on the 
left panel of Fig. 3 by the thick gray curves labeled ``Imaging''.)  
On the other hand, the STCM describes poorly the low momentum 
data for 270 MeV/c $<P_{tot}<$ 390 MeV/c, which have significant 
contributions from moderately to very long-lived sources. This 
is clearly shown in the right panel of Fig. 3, which compares 
the data, the one-parameter Gaussian analysis (dot-dashed line) 
and the best-fit STCM analysis (thin solid line) for the low 
momentum gate. (Note, the vertical scale in the right panel has 
been magnified to allow more sensitive comparisons; the data, 
the single Gaussian source fit and the numerical inversion fit 
are the same as in the left panel.) Clearly, the best fit STCM 
correlation function is wider than the best fit single Gaussian 
source correlation function; the STCM source radius of $r_{0}=4.1$ 
fm is smaller than the single Gaussian source radius of $r_{0}=5.9$ 
fm, consistent with the trends demonstrated in Fig. 2. The measured 
correlation function is even wider, implying that the true source 
is even smaller than that extracted via the STCM, but the correct 
radius information cannot be extracted with the STCM, however, 
because the shape of the source function is actually not Gaussian. 
One could attempt to address the mixture of fast and slow components
by contructing a model of this mixture [27].
However, we believe that accurate information from correlation functions 
when there are significant contributions from long-lived sources may
be better obtained by avoiding apriori assumptions about the shape 
of the source. 

We do this by numerically inverting the correlation functions 
to construct an image of the emission source as described in 
refs. [22-24].  Following ref. [24], the emission sources $S(r)$ 
for the protons were parameterized over $0\leq{r}\leq{20}$ fm 
by six third-order b-spline polynomials, and the numerical 
inversion of Eq. (2) was achieved via the optimization algorithm 
of ref. [23,24]. 
The thick gray curves in Fig. 3 show the best fits to the experimental 
data with this imaging approach. The agreement is excellent. 

The extracted source functions are shown in Fig. 4. The widths 
of these curves represent the lower and upper limits defined 
by one-sigma error bars of the extracted source functions.  The 
uncertainties, small at $r<7$ fm, become very large at $r{\geq}12$ fm. 
 Better data in the small momentum region ($q<10$ MeV/c) of 
the correlation functions could provide improved constraints 
on the source at large radii where the source shape is mainly 
determined by the time scale of the slow emission component. 
Unfortunately, this part of the correlation function is difficult 
to measure and the precision of the data at $q<10$ MeV/c is 
limited, as shown in the left panel of Fig. 3. 

The shapes of the imaged sources and their integrated 
values at $r\geq$12 fm provide clear evidence for a 
two-component structure consistent with the occurrence 
of emission on two very different time scales: a sharply 
localized central region that can be attributed 
to a fast emission component, and a tail region that can 
be attributed to a slow emission component.  
The half-density radii from the radial source profiles, 
$r_{1/2}$, and the yield ratio $f$ from the central 
($r<3{\cdot}r_{1/2}$) region of the imaged source distributions 
are given in Table I. Consistent with the weak momentum dependence 
of the widths $\Delta{q}_{FWHM}$ of the measured correlation 
functions and contrary to the single Gaussian source fits, the 
source radii vary little with $P_{tot}$. 
The strong total momentum dependence of the maximum values of 
the measured correlation functions reflects the momentum dependence 
of the relative contributions from the fast dynamical and slow 
evaporative emitting sources. The fast components dominate the 
high total momentum gate ($f\approx{0.78}$) and become less important at 
lower momenta where they are reduced to about $f\approx{0.30}$ 
for $P_{tot}$=270-390 MeV/c. At such low momenta, long-lifetime 
emissions become more important. This ability of imaging analyses 
to differentiate between fast and slow emissions will enable more 
quantitative comparisons with transport theories in the future. 

In summary, we have investigated the characteristics 
of angle-integrated two-proton correlation functions for scenarios 
in which protons are emitted on two very different time scales. 
 An example of such a scenario would be fast dynamical emission 
from the initial overlap of projectile and target and slow equilibrium 
evaporation or secondary decays of particle unstable fragments. 
When both fast and slow emissions are present, determination 
of the height and width of the correlation peak allow the extraction 
of the initial source geometry associated with the fast component 
and the relative yields from the slow and fast emission processes. 
 More detailed information is obtained by numerically inverting 
the correlation function. Some additional information about the 
lifetime of the slow component might be extracted if the detailed 
shape of the correlation function is measured for very small 
relative momenta.  Precision measurements of angle-integrated 
two-proton correlation functions may thus provide a valuable 
diagnostic tool for reactions where fast dynamical and slow equilibrium 
processes compete. Finally, we note that the simultaneous presence 
of short and long timescale emission may also be relevant for 
two-pion correlations at higher energies where some of the measured 
particles are produced via the decay of hadronic resonances [6,26-29]. 
Whether the imaging techniques so useful for these analyses of 
two-proton correlations can be also useful for two-pion correlations 
is a subject for future investigations.

This work was supported by the National Science 
Foundation under Grant Nos. PHY-95-28844 and PHY-0070818. Part 
of this work was performed under the auspices of the U.S. Department 
of Energy by University of California, Lawrence Livermore National 
Laboratory under Contract W-7405-Eng-48.

\newpage

\textbf{References}

1. S.E. Koonin, Phys. Lett. B 70, 43 (1977). 

2. M. A. Lisa et al., Phys. Rev. Lett. 84, 2798 (2000). 

3. S. Pratt and M.B. Tsang, Phys. Rev. C 36, 2390 (1987). 

4. D.H. Boal, C.K. Gelbke, and B.K. Jennings, Rev. Mod. Phys. 62, 
553 (1990), and references therein. 

5. W. Bauer, C.K. Gelbke, and S. Pratt, Ann. Rev. Nucl. Part. Sci, 
42, 77 (1992), and references therein.

6. U. Heinz and B.V. Jacak, Annu. Rev. Nucl. Part. Sci. 49, 529 
(1999)

7. S. Fritz et al., Phys. Lett. B 461, 315 (1999).

8. H.A. Gustafsson, et al., Phys. Rev. Lett. 53, 544 (1984).

9. W.G. Lynch et al., Phys. Rev. Lett. 51, 1850 (1983). 

10. W.G. Gong et al., Phys. Rev. C 43, 781 (1991).

11. W.G. Gong et al., Phys. Lett. B 246, 21 (1990).

12. W.G. Gong et al., Phys. Rev. C 43, 1804 (1991), and references 
therein.

13. M.A. Lisa et al., Phys. Rev. Lett. 70, 3709 (1993).

14. M.A. Lisa et al., Phys. Rev. Lett. 71, 2863 (1993).

15. M.A. Lisa, Phys. Rev. C 49, 2788 (1994).

16. D.O. Handzy et al., Phys. Rev. C 50, 858 (1994).

17. G.J. Kunde et al., Phys. Rev. Lett. 70, 2545 (1993).

18. S.J. Gaff et al., Phys. Rev. C 52, 2782 (1995).

19. D.O. Handzy et al., Phys. Rev. Lett. 75, 2916 (1995).

20. A. Elmaani et al., Phys. Rev. C 49, 284 (1994).

21. F. Zhu et al., Phys. Rev. C 44, R582 (1991).

22. D.A. Brown and P. Danielewicz, Phys. Lett. B 398, 252 (1997). 

23. D.A. Brown and P. Danielewicz, Phys. Rev. C 57, 2474 (1998). 

24. D.A. Brown and P. Danielewicz, Phys. Rev. C 64, 014902 (2001). 

25.C.J. Gelderloos et al., Phys. Rev. Lett. 75, 3082 (1995).

26. This schematic model recalls the core+halo model introduced 
to describe two-pion correlation functions [4,28,29] There, values 
for $\lambda$ of less than unity have been interpreted as due to coherent 
pion emission or to contributions from long-lived resonance decays 
[4,28].

27. R. Ghetti et al., Nucl. Phys. A 674, 277 (2000). 

28. J.P. Sullivan et al., Phys. Rev. Lett. 70, 3000 (1993)

29. S. Nickerson, T. Cs\"{o}rg\~{o} and D. Kiang, Phys. Rev. C 57, 
3251 (1998)

\newpage

\textbf{Table I}: Values for the half radius $r_{1/2}$ and the fraction 
$f$ of proton emission from the fast component that have been extracted 
following the procedures outlined in the text. The label ``Gaussian'' 
designates the fits using a one-component Gaussian source of 
Eq. (4). (For the Gaussian profiles, $r_{1/2}{\approx}1.18r_{0}$.)

\begin{center}
\begin{tabular}{|c|c|c|c|c|c|c|} 
\hline
\noindent $P_{tot}$ region & \multicolumn{2}{c|}{270-390 MeV/c} &
\multicolumn{2}{c|}{450-780 MeV/c} & \multicolumn{2}{c|}{840-1230
  MeV/c}\\ \hline
             & r$_{1/2}$ (fm) & f & r$_{1/2}$ (fm) & f & r$_{1/2}$
             (fm) & f \\ \hline
Numerical inversion & 2.44$\pm$0.37 & 0.30$\pm$0.05 & 3.13$\pm$0.14 &
0.68$\pm$0.03 & 2.93$\pm$0.15 & 0.78$\pm$0.05 \\ \hline 
Gaussian & 7 & 1 & 5 & 1 & 4 & 1 \\ \hline 
\end{tabular}
\end{center}

\newpage

\textbf{Figure Captions}

\textbf{Figure 1}: The solid line indicates a representative two-proton 
source function for intermediate energy collisions. The dashed 
and dot-dashed lines provide a decomposition of the solid line 
into short and long time scale emissions respectively. Right 
panel: The solid, dashed and dot-dashed curves show the correlation 
functions corresponding to the source functions in the left panel.

\textbf{Figure 2}:  Two proton correlation functions predicted with the 
schematic two-component model (STCM) discussed in the text.  
The left panel gives examples of correlation functions with similar 
heights of the correlation peak, but different widths.  The top 
right panel illustrates the unique relation between source size 
and width of the correlation peak; the line corresponds to a 
linear best fit to the points.  The bottom right panel illustrates 
the relation between the height of the correlation peak, source 
radius and fraction of proton emission originated from fast processes.

\textbf{Figure 3}:  The left panel shows two-proton correlation
functions measured [12] at $\theta_{lab}\approx{25}^{o}$
for $^{14}$N+$^{197}$Au collisions at E/A = 75 MeV for three different 
cuts on the total momentum, 
$P_{tot} = \left|{\mathbf{p}}_{1}+{\mathbf{p}}_{2}\right|$, 
of the coincident protons pairs, 270 MeV/c $<P_{tot}<$ 390 MeV/c 
(circular points), between 450 MeV/c $<P_{tot}<$ 780 MeV/c 
(triangular points), and between 840 MeV/c $<P_{tot}<$ 1230 MeV/c 
(square points). The right panel show the correlation function for 
the 270 MeV/c $<P_{tot}<$ 390 MeV/c cut on an expanded vertical scale. 
The various curves are discussed in the text. 

\textbf{Figure 4}: The solid bands show sources reconstructed from the 
data (Fig. 3) via the inversion technique of refs. [22-24].  
The dot-dashed lines indicate the sources corresponding to the 
Gaussian source fits in the left panel of this figure. For both 
the solid bands and dot-dashed lines, the sources for the highest 
(lowest) momentum gates have the largest (smallest) values at 
$r<2$ fm.

\end{document}